\documentclass[conference,compsoc]{IEEEtran}

\usepackage{amsmath,amsthm}
\usepackage{xcolor}
\usepackage{amssymb}
\usepackage{multirow}
\usepackage{booktabs}
\usepackage{comment}
\usepackage{threeparttable}
\usepackage{subfigure}
\usepackage{pgfplots}
\usepackage{makecell}

\usepackage{colortbl}
\usepackage{float}

\theoremstyle{plain}

\newtheorem{defi}{Definition}

\usepackage{subcaption}
\usepackage{pgfplots}   

\usepackage{bbding}
\usepackage{pifont}
\usepackage{multicol}
\usepackage{tikz}
\usepackage{pgfplots}
\pgfplotsset{compat=1.18}
\usetikzlibrary{patterns}

\usepackage{soul}
\usepackage[linesnumbered,ruled,vlined]{algorithm2e}
\usepackage{algpseudocode}

\usepackage{url}

\newenvironment{packeditemize}{
	\begin{list}{$\bullet$}{
			\setlength{\labelwidth}{4pt}
			\setlength{\itemsep}{0pt}
			\setlength{\leftmargin}{\labelwidth}
			\addtolength{\leftmargin}{\labelsep}
			\setlength{\parindent}{0pt}
			\setlength{\listparindent}{\parindent}
			\setlength{\parsep}{0pt}
			\setlength{\topsep}{1pt}}}{\end{list}}

\newenvironment{takeaway}{
\MakeFramed{\advance\hsize-\width\FrameRestore}}
{\endMakeFramed}

\usepackage{framed}
\definecolor{formalshade}{rgb}{0.95,0.95,0.97}
\definecolor{darkblue}{rgb}{0.14,0.22,0.52}

\usepackage{hyperref}
\ifCLASSOPTIONcompsoc
  \usepackage[nocompress]{cite}
\else
  \usepackage{cite}
\fi

\ifCLASSINFOpdf
\else
\fi
\begin{document}

\title{How To Cook The Fragmented Rug Pull?}

\newcommand{\acro}{{SLID}~}
\newcommand{\nasrin}[1]{\textcolor{red}{#1}}
\newcommand{\TODO}[1]{\textcolor{blue}{#1}}
\newcommand{\minh}[1]{\textcolor{blue}{#1}}
\newcommand{\minhq}[1]{\textcolor{orange}{#1}}
\newcommand{\qw}[1]{\textcolor{magenta}{#1}}
\newcommand{\jas}[1]{\textcolor{blue}{{#1}}}
\newcommand{\bk}[1]{\textcolor{purple}{{#1}}}

% \author{Anonymous Author(s)}

 \author{
  {\rm Minh Trung Tran$^\dag$, Nasrin Sohrabi$^\ddag$, Zahir Tari$^\dag$, Qin Wang$^\S$}\\
  $^\dag$RMIT University $|$ $^\ddag$Deakin University $|$ $^\S$CSIRO's Data61, Australia
%  \and
 % {\rm Nasrin Sohrabi}\\ 
%  Deakin University
 % \and
 % {\rm Zahir Tari}\\
 % RMIT University
 % \and
 % {\rm Qin Wang}\\
 % {\rm CSIRO}
 % \and
 % {\rm Minhui Xue} \\
 % {\rm CSIRO}

  }

\maketitle

\begin{abstract}
Existing rug pull detectors assume a simple workflow: the deployer keeps liquidity pool (LP) tokens and performs one or a few large sells (within a day) that collapse the pool and cash out. In practice, however, many real-world exits violate these assumptions by \textit{splitting} the attack across both time and actor dimensions: attackers break total extraction into many low-impact trades and route proceeds through multiple non-owner addresses, producing low-visibility drains.

We formalize this family of attacks as the \textit{fragmented rug pull} (FRP) and offer a compact “recipe” for a slow-stewed beef special: (i) keep the lid on (to preserve LP control so on-chain extraction remains feasible); (ii) chop thin slices (to split the total exit volume into many low-impact micro-trades that individually fall below impact thresholds); and (iii) pass the ladle (to delegate sells across multiple wallets so that each participant takes a small share of the extraction). Technically, we define three atomic predicate groups and show that their orthogonal combinations yield evasive strategies overlooked by prior heuristics (USENIX Sec’19, USENIX Sec'23).

We validate the model with large-scale measurements. Our corpus contains 303,614 deployed LPs (block ranging from Nov. 2, 2018 to Mar. 5, 2025), among which 105,434 are labeled as FRP pools. The labeled subset includes 34,192,767 pool-related transactions and 401,838 inflated-seller wallets, involving 1,501,408 unique interacting addresses. Notably, owner-wallet participation in inflated selling among FRP-flagged liquidity pools has declined substantially ($\approx$33.1\% of cases), indicating a shift in scam behavior: the liquidity drain is no longer held on the owner wallet. Additionally, we detected 127,252 wallets acting as serial scammers when repeatedly engaging in inflated selling across multiple FRP LPs. Our empirical findings demonstrate that the evasive strategies we define are widespread and operationally significant.

% We validate the model with large-scale measurements. Our corpus contains 303,614 deployed LPs (block range to Mar 5, 2025), of which 105,434 are labeled as FRP-type pools; the labeled dataset comprises 34,192,767 pool-related transactions and 401,838 distinct inflated-seller wallets, covering 1,501,408 unique interacting addresses. Owner wallet participation in inflated selling in FRP-flagged LPs has fallen markedly (owner-involved pools $\approx$ 33.1\%), highlighting the shifting trend of scam where the liquidity drain is no longer held on the owner wallet. Besides, we detected 127,252 different wallets representing serial scammer - when spotted their appearance in conducting inflated selling across multiple FRP LPs. Our empirical findings demonstrate that the evasive strategies we define are quite common.

\end{abstract}

% creates the second title. It will be ignored for other modes.
\IEEEpeerreviewmaketitle

\section{Introduction}~\label{sec:introduction}
Ethereum underpins the majority of decentralized finance (DeFi)~\cite{sokdefi} activity, with approximately 70\% of the total value locked (TVL) across all blockchains~\cite{llama1}. Rug pull scams~\cite{rugpull} remain among the most destructive attack vectors in this ecosystem, leading to annual losses amounting to more than \$2B USD~\cite{rplost}.

A typical rug pull follows a simple pattern. \textit{(i)} The deployer first launches a new token and establishes a liquidity pool on a decentralized exchange, usually pairing the token with a well-recognized base asset such as ETH or USDT to create an illusion of legitimacy and attract early liquidity.  \textit{(ii)} They then supply an initial amount of both assets to bootstrap trading activity and adjust pool parameters to appear stable and liquid.  Once the pool is live, the deployer engages in aggressive marketing through social media announcements, airdrops, and influencer endorsements to stimulate speculative demand and drive token purchases from unsuspecting investors. As more participants join, the token’s market valueincreasese, masking the asymmetry of control retained by the deployer over the liquidity provider (LP) tokens. \textit{(iii)} At the moment the liquidity peak, the attacker executes the exit phase: selling large quantities of the project token or directly withdrawing the LP tokens to reclaim the base asset. The sudden extraction (usually within one day) drains the pool’s reserves, causes the token’s price to collapse, and leaves investors with unsellable assets, marking the completion of the rug pull.

%the deployer creates a liquidity pool, adds an initial token pair, promotes it to attract investors, and later dumps their holdings for the base asset, causing the pool’s price and liquidity to collapse.

% \begin{figure}[t]
% \subfigure[\textbf{Traditional rug pull heuristic} detection relies on three conditions (i.e., Predicates A-C). If all three hold, the pool is flagged as a rug pull.  However, we find that B and C are \emph{modifiable} and can be further generalized! ]{\label{fig:tradi}
% \begin{minipage}[b]{\linewidth}
% \centering
% \includegraphics[width=1\linewidth]{graphs/tradi.png}
% \end{minipage}
% }
% \vspace{5pt}
% \subfigure[\textbf{Our extended FRP heuristic} generalizes (B) and (C) into \emph{dynamic, multi-actor observables} to
% capture both \emph{impact fragmentation} (splitting exits into low-impact fragments) and \emph{identity delegation} (distributing sells across multiple non-owner addresses). These extensions can bypass existing static heuristics and motivate our on-chain measurement of predicate-level evasions.
% ]{ \label{fig:frgIntro}
% \begin{minipage}[b]{\linewidth}
% \centering
% \includegraphics[width=1\linewidth]{graphs/ours.png}
% \end{minipage}
% }

% \caption{Fragmented Rug Pull}
% \label{fig-howtransfer}
% \vspace{-0.2in}
% \end{figure}

\subsection{Why Existing Detection Fails?}

\begin{figure}[t]
\subfigure[\textbf{Traditional rug pull heuristic} detection relies on three conditions (i.e., Predicates A-C). If all three hold, the pool is flagged as a rug pull.  However, we find that B and C are \emph{modifiable} and can be further generalized! ]{\label{fig:tradi}
\begin{minipage}[b]{\linewidth}
\centering
\includegraphics[width=1\linewidth]{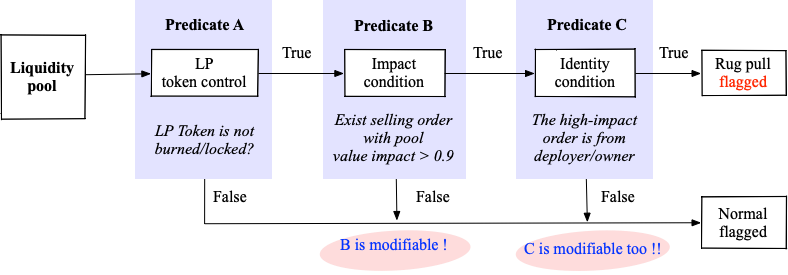}
\end{minipage}
}
\vspace{5pt}
\subfigure[\textbf{Our extended FRP heuristic} generalizes (B) and (C) into \emph{dynamic, multi-actor observables} to
capture both \emph{impact fragmentation} (splitting exits into low-impact fragments) and \emph{identity delegation} (distributing sells across multiple non-owner addresses). These extensions can bypass existing static heuristics and motivate our on-chain measurement of predicate-level evasions.
]{ \label{fig:frgIntro}
\begin{minipage}[b]{\linewidth}
\centering
\includegraphics[width=1\linewidth]{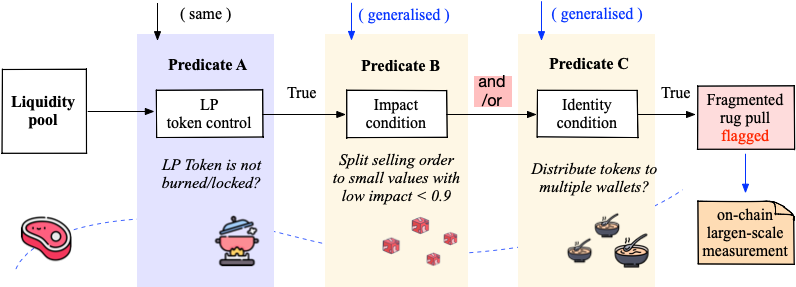}
\end{minipage}
}
\vspace{-0.1in}
\caption{Fragmented Rug Pull}
\label{fig-howtransfer}
\vspace{-0.1in}
\end{figure}

Early measurement and forensic studies~\cite{rugpull,spammer} distilled rug-pull behavior into simple operational heuristics (Figure ~\ref{fig:tradi}). Concretely, detectors raise alerts when two observable conditions (we denote it as \textit{predictes} to emphasize their role as a family of structurally similar logical elements) co-occur: \textit{(i)} a large sell or withdrawal that sharply reduces pool liquidity (impact), and \textit{(ii)} the transaction’s sender is the token owner or LP deployer (identity).  
These impact- and identity-based rules underpin both academic and commercial detectors and have long served as effective first-order signals for classic, one-shot exits.

However, attackers adapted faster than the heuristics. 

Our reanalysis of on-chain incidents since 2021 (\S\ref{sec:motivating-examples}) shows a clear shift in operational patterns: \textit{instead of a single, high-impact owner sell, adversaries increasingly split extraction into many smaller trades and route them through non-owner addresses. }
The aggregate economic effect is the same, yet no single transaction violates the legacy per-transaction checks; consequently, many confirmed scams no longer exhibit the textbook \textit{single large owner sell} signature relied upon by rule-based detectors. 
This structural divergence produces a blind spot: pools are drained economically like rug pulls, but remain invisible to static heuristics.

The root cause is fundamental: existing heuristics encode an observable \emph{symptom} (a conspicuous transaction) rather than the underlying \emph{economic gap} (that the attacker can extract value).  
Once adversaries know the fixed triggers, evasion is straightforward: change who sells, reduce per-transaction volume, or alter timing.

A recent study~\cite{slid} extended detection into the temporal dimension, reporting the emergence of \emph{slow liquidity drain} (SLID) scams where attackers siphon funds gradually rather than through a single abrupt exit.  
While SLID effectively lowered the detection threshold by monitoring liquidity loss over extended periods, its analysis remains only temporal, without revealing how attackers alter the underlying detection logic or predicate conditions.

What is still missing is a principled understanding of which predicate-level modifications preserve the attacker’s payoff, and how frequently such evasive variants appear in the wild. Addressing these gaps motivates our formal decomposition of rug-pull heuristics into atomic predicates.

\subsection{Our Efforts}

%Motivated by LIBRA (2025), Safereum (2023), ZhongHua (2024), and 20 additional examples \S \ref{sec:motivating-examples}, we formalize a new family of evasive exits:

We progress by identifying and addressing a series of research questions (RQs):

\smallskip  
\begin{packeditemize}
  \item \textbf{RQ1:} \textit{Which canonical heuristic components of rug pulls are brittle under adversarial modification?}
\end{packeditemize}
\smallskip

Traditional rug-pull detection relies on a small set of hand-crafted behavioral rules to flag high-impact, owner-originated liquidity exits.  
However, prior studies lack a formal decomposition of these heuristics into elementary predicates and overlook which parts can be adversarially modified without reducing the attacker’s economic return.

To fill this gap, we revisit canonical detection logic and express it as a minimal set of measurable predicates covering three behavioral dimensions:  
(A) \emph{liquidity-pool control},  
(B) \emph{splitting large exits} (keeping each order’s impact below a threshold), and  
(C) \emph{identity obfuscation via multi-wallet execution}.  
Through theoretical analysis, we identify which predicates can be modified without lowering attacker payoff, providing the first principled proof of heuristic brittleness.

This predicate-level view reveals two core evasive operations aligned with (B) and (C):  
(i) \emph{splitting large exits} into many low-impact orders, and  
(ii) \emph{delegating execution across multiple wallets} to conceal ownership.  

A complementary tactic is to \emph{schedule trades over time} (temporal smoothing, as explored in ~\cite{slid}) to further dilute observable signals, though this acts as an overlay on (B)/(C) rather than a standalone predicate.

\smallskip  
\begin{packeditemize}
  \item \textbf{RQ2:} \textit{How can these evasive modifications be formally modeled to define and generalize a new scam type?}
\end{packeditemize}
\smallskip  

Building on the above predicate analysis, we formalize a new class of evasive scams and, termed the \emph{Fragmented Rug Pull (FRP)}, to systematically exploit heuristic brittleness.

\begin{defi}[\textbf{Fragmented rug pull (FRP)}, informal sketch]
A fragmented rug pull is an exit strategy in which the attacker preserves liquidity-pool control (Predicate~A) while extracting value by  
(i) \emph{splitting large exits} so that each order remains below the impact threshold (Predicate~B), and  
(ii) \emph{obfuscating identity} by delegating execution across multiple wallets (Predicate~C),  
ensuring that no single transaction or address triggers canonical rug-pull heuristics.
\end{defi}

Formally, FRP behavior can be characterized by three independent parameters:  
(i) the number of selling wallets (\(N\)),  
(ii) the per-wallet order count (\(K_a\)), and  
(iii) the per-order impact (\(v\)),  subject to the invariant that each order’s impact satisfies \(v < \theta\) (the canonical impact threshold) and that ownership signals are dispersed across wallets so that \emph{no single owner} exhibits a heuristic-triggering footprint.  

Our model unifies evasive behaviors under a single analytical framework, making FRPs quantifiable and comparable to traditional rug pulls. We show more in \S\ref{sec:proving}.

\smallskip  
\begin{packeditemize}
  \item \textbf{RQ3:} \textit{How prevalent are these fragmented rug pulls in real-world DEX environments?}
\end{packeditemize}
\smallskip  

To answer this, we perform a large-scale on-chain measurement covering six major Ethereum DEXs (Uniswap, SushiSwap, Balancer, Curve, PancakeSwap, and BancorSwap) from November~2018 to March~2025.  
Among 384{,}029 liquidity pools and over 1.1~billion transactions, 303{,}614 short-lived pools (lifetime~$\leq$~100~days) were retained for analysis.  

Applying our predicate-based FRP definition, we identify 105{,}434 pools (34.7\%) as fragmented rug pulls, involving 34.19M transactions, 401{,}838 inflated-seller wallets, and 1.50M unique interacting addresses.

We find that fragmented rug pulls have rapidly become a mainstream scam pattern in DeFi.  
Their presence was negligible before 2020 (fewer than one hundred cases were observed) but grew explosively with the rise of decentralized trading, reaching nearly 40k detected pools in 2023.  

Our measurement further shows a clear shift in how attackers operate. 
Traditional single‐wallet exits are steadily disappearing (from \textbf{57\%} in 2019 to \textbf{28.3\%} in 2024), while multi‐wallet schemes have become the norm—over 70\% of FRPs now involve coordinated actions from a few sellers rather than one.  
Owner participation has dropped from \textbf{65\%} to \textbf{24\%}, showing that deployers increasingly rely on delegated wallets to hide their identity.  
Single‐wallet scams finish within a couple of days and involve only a handful of sells, whereas multi‐wallet schemes stretch over a week with dozens of small trades that mimic normal market activity.

\smallskip
Our contributions, in short, are

\begin{itemize}
  \vspace{-0.1in}
  \item predicate decomposition of rug pulls;
  \item new definition of fragmented rug pull scam;
  \item formal detection model as extended heuristics; 
  \item large-scale on-chain measurement within Ethereum.
\end{itemize}

Beyond modeling and measurement, we further \textbf{constructed and released}\footnote{Accessible via \url{https://anonymous.4open.science/r/FRP_source-730C}} our labeled pool dataset~\cite{labeledfrpdataset}, predicate evaluations, and analysis scripts to support reproducibility and future comparative studies~\cite{frpsource}.  
  
We also summarized practical indicators that can serve as early warning signals for emerging scams, such as the creation of multiple distributor wallets shortly after deployment, synchronized low‐impact sell patterns, and rapid post‐exit fund consolidation. We hope that our open resources can facilitate the future on‐chain fraud detectors that reason over aggregated behaviors rather than isolated events.

\section{Technical Warm-Ups}\label{sec:bck}

\subsection{Ethereum Token and Account model}
\label{sec:ethereum}

We review three technical facts that enable fragmented exit strategies: 
(i) ERC-20's interface permits programmatic transfer of value across heterogeneous addresses;  
(ii) approval and minting / pre-funding functions make bulk distribution and inflated selling operationally feasible; and  
(iii) the combined use of EOAs and contract yields multi-address flows that evade single-transaction heuristics.

\smallskip
\noindent\textbf{ERC-20.}
Ethereum's fungible token implements the ERC-20 standard~\cite{erc20}, which defines a uniform transfer and approval APIs and emits \textsf{transfer} events. This improves composability across DEXs, bridges, and lending protocols, but the identical call patterns can be leveraged by attackers.

\textsf{approve}/\textsf{transferFrom} enables pre-authorised token transfer by third parties or contracts, enabling automated distribution from one source to many intermediary addresses. Token contracts that allow \textsf{mint} make pre-funding and inflated balances possible, supporting large downstream sells without immediate owner-originated buys.

\smallskip
\noindent\textbf{EOAs, contracts, and internal transfers.}
Ethereum distinguishes Externally Owned Accounts (EOAs) and Contract Accounts~\cite{wang2023account,ethaccounttype}. Contracts can programmatically split, route, and synchronize transfers (including multi-hop calls), producing multiple internal transfers under a single external \texttt{tx} hash. The result is multi-address fund flows that are difficult to flag using per-transaction heuristics.

\subsection{Decentralized Exchanges and Liquidity Pools}
\label{sec:dex}

\noindent\textbf{Decentralized exchanges.} DEXs form the backbone of the DeFi ecosystem, enabling users to trade assets directly from their wallets. Uniswap~\cite{uniswap0}, the largest DEX on Ethereum, exemplifies this model. Unlike centralized exchanges that hold user assets, DEXs execute trades on-chain through automated market makers (AMMs)~\cite{dexamm}. Each AMM governs a liquidity pool (LP). LP is a reserve of token pairs priced continuously by an invariant function. LP allows permissionless trading without intermediaries.

\smallskip
\noindent\textbf{Liquidity pool.}
An LP typically holds a base token (e.g., ETH, USDT, USDC, DAI) and a paired token, often a newly issued asset. 
The pool creator deposits both to initialize the price. Most DEXs adopt the constant-product model~\cite{ammformula,uniswapdexpriceexplain}, \(x \cdot y = k\), where \(x\) and \(y\) denote token reserves and \(k\) is constant. This design ensures continuous liquidity but induces price slippage: larger trades cause proportionally higher price impact.

Users interact with LPs through deposits, withdrawals, and swaps, all recorded on-chain. 
Buying the paired token increases the base-token reserve and raises its price; selling does the opposite.  Any participant may also provide liquidity by depositing both tokens, receiving LP tokens as proof of ownership and later redeeming them for their share.

Control over LP tokens determines control over the pool’s liquidity.  If deployers retain these tokens, they can later withdraw reserves.  Legitimate projects burn or lock LP tokens~\cite{sokdefi,jiang2023decentralized,uniswap0,lpburn} to prevent such withdrawals and signal credibility. In contrast, scammers typically avoid or falsify this step, preserving withdrawal rights while disguising future liquidity drains as normal trading activity.

\subsection{Withdraw vs. Selling in Rug Pull}
\label{subsec:rugpull-lit}

A rug pull is an LP-based exit scam in which an attacker converts pool-held value into base assets, leaving the remaining investors with devalued tokens~\cite{rugpull,spammer}.

In practice, rug pull scammers employ two economically equivalent extraction preferences:

\begin{packeditemize}
  \item \textit{Withdraw.} The deployer redeems unburned LP tokens to remove liquidity directly from the pool after users add value by buying the paired token. A single withdrawal can drain most of the base-token reserves in one atomic action, triggering an abrupt price collapse and producing an unmistakable owner-originated trace on-chain.
  \item \textit{(Inflated) selling.} The attacker sells an inflated supply of the paired token into the pool to extract base tokens indirectly. Large, concentrated sells deplete reserves through price slippage, while repeated or coordinated sells can achieve the same depletion effect as a direct withdrawal.
\end{packeditemize}

When the total value recovered from withdrawals or inflated sells exceeds the attacker’s initial stake (after fees and slippage~\cite{rugpull,spammer,slid}), the attacker gains a profit.
Heuristic detectors identify rug pulls by flagging abrupt owner withdrawals or high-impact sell actions~\cite{xu2019anatomy,la2023doge}.

Throughout this paper, \textit{selling} by scam wallets refers to this \textit{inflated selling}, not routine trading behavior.

% \section{Motivational Examples}\label{sec:motiv}

\section{Motivational Examples}
\label{sec:motivating-examples}

We revisit concrete on-chain incidents. Each case results in a similar outcome: a near-total liquidity collapse. But they exit in different ways and evade automated flagging. 

\smallskip
\noindent\textbf{Case study on LIBRA.}
LIBRA is one of the biggest cryptocurrency scams~\cite{libra} so far. It reached over \$5 billion in token value just a few hours after launch.

The deployer retained control of the initial liquidity tokens yet performed no large or visible owner sells. Instead, the exit happened through four new wallets created within one hour of deployment. Each wallet made only a few medium-sized (two - five) sells. Every sell stayed below the usual detection limit (\(\theta\approx0.9\)) used by prior tools~\cite{rugpull,spammer}. Small buy orders and private-bundled trades were mixed in to hide price changes.

LIBRA caused the same results as a regular rug pull and avoided triggering any known rug pull heuristics. No owner transaction exceeded the impact threshold and per-transaction impact metrics remained within “normal” volatility ranges.  Automated detectors marked the pool as normal, and the scam went unnoticed for days. LIBRA was ultimately identified only through secondary signals (off-chain alerts and owner self-confession on intentional scam purpose) rather than the canonical on-chain heuristics. 

\smallskip
\noindent\textbf{Additional cases.}
Safereum~\cite{safereum} and ZhongHua~\cite{zhonghua} used similar evasion strategies with slightly different methods. Safereum deployed about a dozen pre-funded wallets. Each wallet made a short series of small sells mixed with normal-looking trades. The profits were then sent to one shared wallet and later bridged out. ZhongHua took a different approach. It used a proxy contract to make timed sells through delegated EOAs while the main deployer stayed inactive, giving the illusion of independent traders. Both cases were later confirmed as scams by independent on-chain investigation platforms and community security trackers. Specifically, Safereum was flagged by multiple analytics aggregators for coordinated multi-wallet drains by Certik, while ZhongHua was reported through cross-chain tracing that linked one of the inflated selling addresses to another group of identified scam wallets. %(more details in Appendix~\ref{appendix:motivationother}).

Beyond these two, we analyzed more liquidity pools showing the same pattern: fragmented exits, multiple selling wallets, and no large sells from the owner. Each case produced the same invisible outcome as LIBRA (details in~\cite{moremotiveex}).

In all examples, no single sell exceeded the detection threshold or came directly from the deployer’s wallet. As a result, existing automated detectors missed all of them. 
\section{Formalizing Fragmented Rug Pull}\label{sec:proving}

Our motivational examples show that existing detectors can be evaded in practice. To investigate this gap, we decompose the canonical rug-pull heuristic into elemental predicates and analyze which assumptions an attacker can change without altering the scam’s economic payoff (e.g., LP control, per-transaction impact, seller identity).

We proceed in two coordinated steps:
\begin{packeditemize}
  \item \textit{Formal decomposition.} We formalize the baseline sell-based heuristic as a small set of measurable atomic predicates and express each predicate mathematically so its dependence on per-transaction magnitude, actor identity, and timing is explicit. Building on this formalization, we specify an adversary model and a family of predicate-level modifications (an attacker can apply, e.g., fragmenting impact, delegating identity) that preserve the economic outcome while aiming to evade detection.
  \item \textit{Empirical evaluation.} We apply the formal FRP definition to on-chain traces to label affected pools and quantify predicate-level evasions. Our measurement includes (i) actor-centric analysis (who performs drains: wallet distribution, owner participation, recurrence) and (ii) action-centric analysis (how drains are executed: sell frequency, fragmentation level, behavioral categories). 
\end{packeditemize}

\subsection{Sell-Based Rug Pull Predicates}
\label{subsec:heuristic-formalization}

Existing detectors~\cite{rugpull,spammer} identify rug pulls by searching for large owner sells, LP burns/unlocks, and sharp liquidity drawdowns as key behavioral signatures (demonstrated in Figure~\ref{fig:tradi}). 
We take this baseline as the foundation for formalization.

\smallskip
\noindent\textbf{Notation.}
Let $\mathcal{P}$ denote a DEX liquidity pool, and let $T=\{t_1,\dots,t_n\}$ be the time-ordered set of \emph{sell/withdraw} transactions that affect $\mathcal{P}$.
For each transaction $t_i$ where $t_i\in T$, we define:
\begin{packeditemize}
  \item $a_i$: the \emph{effective actor} that executes the pool-affecting action in $t_i$ (the address initiating the swap or LP removal);
  \item $\tau_i$: the timestamp of $t_i$;
  \item $s_i$: the paired-token amount sold or withdrawn in $t_i$;
  \item $v_i$: the realized base-token value returned by the DEX to $a_i$ in $t_i$;
  \item $\mathcal{V}_i$: the base-token value of the pool before $t_i$.
\end{packeditemize}

\smallskip
\noindent\textbf{Owner set.}
Let $O$ denote the deployer or LP creator address.  
Define $\mathcal{O}$ as the \emph{owner set} used by baseline heuristics.  
In its conservative form, $\mathcal{O}=\{O\}$; it may be expanded to include proxy or custodian addresses when such mappings are identifiable.

\smallskip
\noindent\textbf{Atomic predicates.}
We next formalize the basic conditions (predicates) used in heuristic detection.  
These predicates describe, respectively, LP ownership retention, transaction impact magnitude, and ownership identity of the actor.
\begin{align*}
\mathrm{Impact}(t_i) &\triangleq \frac{v_i}{\mathcal{V}_i}, \\
\mathrm{SellerIsOwner}(t_i) &\triangleq (a_i \in \mathcal{O}), \\
\mathrm{RetainLP} &\triangleq \neg \mathrm{burnLock}(\mathrm{LPT}_{\mathrm{init}}),
\end{align*}
where $\mathrm{LPT}_{\mathrm{init}}$ denotes the initial LP tokens controlled by the deployer, and $\mathrm{burnLock}(\cdot)$ returns true if those tokens were irreversibly burned or time-locked.

We define the predicates as below.

\begin{align}
&\mathrm{RetainLP}. \tag{A}\label{eq:A}\\
&\exists\, t_i\in T:\; \mathrm{Impact}(t_i)>\theta. \tag{B}\label{eq:B}\\
&\mathrm{SellerIsOwner}(t_i). \tag{C}\label{eq:C}
\end{align}

Here, $\theta \in (0,1)$ denotes the threshold for a single-transaction impact, commonly set to $\theta=0.9$ following prior studies~\cite{rugpull,spammer}.

\smallskip
\noindent\textbf{Detection rule.}
The detector $\mathcal{D}$ flags $\mathcal{P}$ as a rug pull if:
\[
  \mathcal{D}(\mathcal{P}) = \eqref{eq:A} \wedge \eqref{eq:B} \wedge \eqref{eq:C}.
\]

Predicate~\eqref{eq:A} ensures the deployer retains LP control;  
\eqref{eq:B} captures a large one-shot sell or withdrawal that significantly depletes liquidity;  
and \eqref{eq:C} links that action to the owner set.  

An attacker can evade $\mathcal{D}$ by violating any one of these predicates:
(i) concealing LP control (e.g., by locking or redistributing LP tokens),  
(ii) fragmenting exits so that no individual $\mathrm{Impact}(t_i)$ exceeds $\theta$, or  
(iii) delegating draining transactions to non-owner addresses ($a_i \notin \mathcal{O}$).  
Such predicate-level manipulations preserve the same economic outcome while escaping heuristic detection.

Notably, in our empirical evaluation (\S\ref{subsec:inflated-categories}), we further vary $\theta$ from 0.7 to 0.95 to assess the sensitivity of the baseline detector under different impact thresholds.

%====================

\subsection{Applying Modifications to Adversaries}
\label{subsec:adversary-perturbations}

Having decomposed the canonical heuristic into its atomic predicates~\eqref{eq:A}--\eqref{eq:C}, we now reason formally about how an attacker can perturb these predicates while preserving the same payoff-extraction mechanism as a canonical rug pull.  
This section enumerates feasible predicate-level modifications that can transform an originally detectable rug pull into an operationally equivalent, but heuristically undetectable, variant.

We organize the analysis by predicate:
\begin{packeditemize}
    \item \textit{modifications of Predicate~\eqref{eq:A}}: can LP-token ownership be changed without losing payoff? (see \S\ref{subsec:modify-a})
    \item \textit{modifications of Predicate~\eqref{eq:B}}: how can impact-based rules be bypassed by fragmentation? (see \S\ref{subsec:modify-b})
    \item \textit{modifications of Predicate~\eqref{eq:C}}: what identity obfuscations make large sells appear non-owner originated? (\S\ref{subsec:modify-c})
\end{packeditemize}

%====================

\subsubsection{Retaining LP control is necessary}
\label{subsec:modify-a}

Predicate~\eqref{eq:A} requires the deployer to retain control of the initial LP tokens (i.e., $\mathrm{RetainLP}$). If the attacker irrevocably burns or time-locks these tokens, they forfeit any on-chain capability to withdraw or self-sell liquidity later; hence the canonical extraction route disappears.

Let $\Pi$ denote any extraction plan that violates~\eqref{eq:A} (e.g., $\mathrm{burnLock}(\mathrm{LPT}_{\mathrm{init}})=\text{true}$). Then for every realization of $\Pi$ the attacker’s on-chain return $\mathcal{R}_\Pi$ is strictly smaller than the canonical return $\mathcal{R}_{\mathrm{canonical}}$ (which assumes retained LP control). Formally,
\[
\mathcal{R}_\Pi \;<\; \mathcal{R}_{\mathrm{canonical}}.
\]
Consequently, realistic evasions aiming to produce an equivalent economic outcome must preserve~\eqref{eq:A}; strategic modifications therefore focus on predicates~\eqref{eq:B} and~\eqref{eq:C}.

%====================

\subsubsection{Per-transaction fragmentation}
\label{subsec:modify-b}

Predicate~\eqref{eq:B} triggers an alert when at least one transaction $t_i\!\in\!T$ satisfies
\[
\mathrm{Impact}(t_i) = \frac{v_i}{\mathcal{V}_i} > \theta .
\]
Under canonical assumptions, $\theta\!\approx\!0.9$ represents a high-impact sell or withdrawal that depletes most of the pool’s reserves in a single action~\cite{rugpull,spammer}.  
The attacker can circumvent this predicate by dividing a large extraction into multiple smaller fragments such that each individual trade remains below the detection threshold.

\smallskip
\noindent\textbf{Fragmentation principle.}  
Let the attacker aim to extract a total base-token value $V_{\mathrm{total}}$ from pool~$\mathcal{P}$.  
The attacker chooses a partition of $V_{\mathrm{total}}$ into $K$ fragments $(v_1,\dots,v_K)$ satisfying
\[
\forall j,\quad \frac{v_j}{\mathcal{V}_j} \le \theta.
\]
Then no single transaction violates~\eqref{eq:B}, yet the cumulative proceeds $\sum_{j=1}^K v_j$ can approximate the same aggregate withdrawal as the canonical rug pull.  
Operationally, this strategy requires only that the attacker sequentially execute $K$ transactions rather than one, which is trivial to automate.

\smallskip
\noindent\textbf{Economic feasibility.}  
Fragmentation incurs additional transaction fees and may suffer from compounding slippage.  
Let $\mathrm{slip}(v_j)$ denote the slippage cost for a trade of size~$v_j$, and $G$ the cumulative gas and relay fees across all fragments.  
The attack remains profitable when
\begin{equation}
\sum_{j=1}^{K} \big(v_j - \mathrm{slip}(v_j)\big) - G \ge V_{\min},
\label{eq:feasible}
\end{equation}
where $V_{\min}$ is the attacker’s minimal acceptable return.  
Because automated market maker (AMM) price curves are convex~\cite{ammformula,uniswapdexpriceexplain}, marginal slippage increases superlinearly with trade size; hence breaking a single large transaction into multiple smaller ones can reduce per-trade price impact faster than it raises total cost.  
This ensures that feasible $(K,G)$ pairs exist for many pools, especially when gas fees are low and the attacker controls both sides of the swap.

\smallskip
\noindent\textbf{Profit-satisficing fragmentation.}  
Attackers may not insist on achieving the \emph{exact} canonical payoff.  
Instead, they can adopt a satisficing approach of accepting a smaller but positive profit to greatly reduce exposure.  
Formally, the attacker selects $K$ and $\{v_j\}$ such that each $\mathrm{Impact}(t_j)\!\le\!\theta$ and inequality~\eqref{eq:feasible} holds for some modest $V_{\min}\!>\!0$.  
This approach balances profit and stealth: once net proceeds exceed deployment and transaction costs, further extraction only increases visibility and risk with diminishing marginal gains.

\smallskip
\noindent\textbf{Behavioral rationale.}  
Bounded-rational attackers often “satisfice” rather than maximize profit~\cite{simon1955behavioral}.  
From a deterrence-theoretic view~\cite{becker1968crime}, expected utility declines as detection probability rises; therefore, reducing transaction amplitude while maintaining a positive return is a rational choice.  
In practice, many real-world drains exhibit such fragmentation patterns, which moderate in size, temporally spaced, and collectively sufficient to yield net positive profit after fees and slippage.  
Throughout this paper, we consider a drain \emph{profitable} once the attacker’s total recovered asset value exceeds initial capital and cumulative operational costs, consistent with prior illicit-profit analyses~\cite{slid}.

%====================

\subsubsection{Seller-identity obfuscation}
\label{subsec:modify-c}

Predicate~\eqref{eq:C} attributes a large-impact transaction to the owner set~$\mathcal{O}$ through the predicate
$\mathrm{SellerIsOwner}(t_i) \Leftrightarrow (a_i \in \mathcal{O})$.  
In the canonical heuristic, a pool is flagged when a high-impact transaction
satisfies both $\mathrm{Impact}(t_i)>\theta$ and $a_i\in\mathcal{O}$.  
Therefore, to falsify~\eqref{eq:C}, the attacker must ensure that all impactful sells are executed by addresses outside of $\mathcal{O}$.

\smallskip
\noindent\textbf{Core idea.}  
Identity obfuscation does not require cryptographic anonymity; it only needs to break the on-chain link between the deployer and the seller executing the drain.  
The attacker can route exit transactions through a set of \emph{distributor addresses} $\mathcal{A}=\{a^{(1)},\dots,a^{(N)}\}$, each distinct from any $O\in\mathcal{O}$.  
Formally, the attacker enforces
\[
\forall\, t_i \in T_{\mathrm{exit}},\quad a_i \notin \mathcal{O}.
\]
From the detector’s viewpoint, this immediately falsifies Predicate~\eqref{eq:C}, since no large-impact trade originates from an address known to be the deployer or owner.

\smallskip
\noindent\textbf{Evasion design space.}  
Seller-identity obfuscation can be implemented across three increasing levels of sophistication:

\begin{packeditemize}
    \item \textit{Direct delegation.}  
    The simplest form uses manually controlled EOAs that receive temporary liquidity from the owner, execute the sells, and return proceeds off-chain or through private transfers.  
    These addresses may be newly generated (``fresh'') or pre-funded by the owner through intermediary swaps.

    \item \textit{Proxy or contract-level delegation.}  
    A more robust variant uses intermediate contracts that perform the selling on behalf of the owner, e.g., minimal proxies, flash-loan wrappers, or batch relays.  
    Because the selling contract address differs from $O$, the transaction is not labeled as owner-originated even though the underlying control remains centralized.

    \item \textit{Cross-wallet or cross-chain laundering.}  
    Sophisticated actors combine fragmentation and delegation by splitting liquidity across chains or DEX instances, routing tokens through bridges or mixers before reconvergence.  
    This dilutes both temporal and identity correlations, further reducing the probability that any single address triggers~\eqref{eq:C}.
\end{packeditemize}

\smallskip
\noindent\textbf{Operational feasibility.}  
Such identity-based evasions are low-cost and easily automated.  
Creating or funding new EOAs incurs negligible on-chain cost; gas overhead grows linearly with the number of distributor addresses~$|\mathcal{A}|$, and can be further amortized through batch transactions or relays.  
The main coordination constraint is ensuring that the cumulative exit volume across $\mathcal{A}$ satisfies
\[
\sum_{a^{(k)}\in\mathcal{A}} \sum_{t_i:a_i=a^{(k)}} v_i = V_{\mathrm{total}},
\]
while preserving $\mathrm{Impact}(t_i)\le\theta$ for each fragment.  
When combined with the fragmentation strategy of Predicate~\eqref{eq:B}, these constraints are easily met.

\smallskip
\noindent\textbf{Stealth properties and forensic implications.}  
Identity obfuscation attacks degrade heuristic accuracy in two ways:
(i) they suppress direct owner–transaction associations required by~\eqref{eq:C}, and  
(ii) they increase the entropy of observable addresses, forcing detectors to infer ownership indirectly through behavioral clustering or timing analysis.  
Moreover, when multiple distributors act in parallel, their combined activity mimics organic market selling rather than a single coordinated drain, further reducing detection confidence.

\smallskip
\noindent\textbf{Beyond owner obfuscation.}  
Attackers can also hide ownership semantically, e.g., transferring control to a DAO-like~\cite{fabrega2025voting,wang2025understanding} multisig or locking LP tokens under a timelock contract while simultaneously retaining trading rights through a separate contract.  
In such cases, even static ownership labeling (e.g., ``contract deployed by $O$'') becomes ambiguous, as legitimate DEX operations share similar structures. We exclude such cases within this paper.

\begin{takeaway}
\noindent\textbf{Finding (\S\ref{sec:proving}).}  
Our formal decomposition highlights that canonical rug-pull detectors rely on three brittle predicates: LP control~\eqref{eq:A}, high-impact sells~\eqref{eq:B}, and owner-originated activity~\eqref{eq:C}.  
Among these, only~(A) is immutable for maintaining profitability.  
Predicates~(B) and~(C) can be independently modified without reducing the scam’s payoff.  
By fragmenting exit volume (violating (B)) or delegating sells to non-owner addresses (violating~(C)), attackers can evade heuristic triggers while retaining equivalent or satisficing profit.
\end{takeaway}

\subsection{Fragmented Rug Pull}% - The Definition
\label{sec:def}

Attackers in the wild combine the elementary variants above into a single operational plan. These combinations vary in engineering cost and stealth, but they all exploit the same logical weakness of the canonical detector: the heuristic looks for a small set of atomic signatures (mentioned as \eqref{eq:A}, \eqref{eq:B} and \eqref{eq:C}). By composing perturbations, an attacker can ensure that these 3 atomic checks cannot be simultaneously true while still realizing a profitable extraction.

Practically, an evasive exit is most naturally described by three core, nested degrees of freedom:

\begin{packeditemize}
  \item \textit{Number of seller wallets ($N$)}: the count of distributor/seller wallets that perform the inflated selling activity, can range from 1 (a single distributor) to many.
  \item \textit{Per-wallet selling order count ($K_a$)}: for each seller $a$ among those $N$ wallets, the number of sell orders executed by $a$; each $K_a$ can range from 0 to many (and only 1 sell from any of $N$ is enough for a profitable result).
  \item \textit{Per-order volume ($v$)}: for every sell order in each $K_a$, the order size may range from very small fragments up to amounts approaching (but typically not exceeding) the detector threshold $\theta$, which is 0.9 based on \cite{rugpull}.
\end{packeditemize}

Operationally, the attacker picks $N$ distributor wallets, assigns each a sequence of $K_a$ orders, and chooses per-order volumes $v$ for those orders. Owner participation (whether the deployer address $O$ also sells/participates in the group of inflated selling wallet $N$) is orthogonal to this construction: it is a boolean side-choice that does not change the combinatorial structure above. The simple invariant required to evade owner+impact heuristics is that \textit{no large-impact transaction is attributed to $O$} ($N = 1$ and not owner, $v \geq 0.9$ in any $K_a$), beside that combination, the attacker is free to choose any combination of $N$, $K_a$, and $v$ based on their preference and still be able to result in fraudulent profit.

From the above predicate analysis and how to break them in heuristic (which is visualized in detail in Figure~\ref{fig:frgIntro}), we present the formal definition of FRP scam.

\begin{defi}[\textbf{Fragmented rug pull (FRP)}]
An LP is flagged as a fragmented rug rull (FRP) scam over the LP lifetime of less than 100 days (\S\ref{subsec:tradeoff}) if the following hold:
% \minh{\begin{packeditemize}
%     \item[i.] \textbf{Control retention:} $\mathrm{burnLock}(\mathrm{LPT}_{\mathrm{init}})=\text{false}$. 
%     The deployer maintains full control over the initial LP tokens.
%     \item[ii.] \textbf{Fragmented inflated selling.} 
%     There exists a set of inflated seller(s) address $\mathcal{A}=\{a_1,\dots,a_N\}$ and, for each $a\in\mathcal{A}$, a sequence of sell events 
%     $S_a=\{(t_{a,1},v_{a,1}),\dots,(t_{a,K_a},v_{a,K_a})\}$ such that
%     \[
%     \forall a\in\mathcal{A},\;\forall j\!\in\![K_a]:\quad \mathrm{Impact}(t_{a,j})=\frac{v_{a,j}}{\mathcal{V}(t_{a,j}^-)} \le \theta,
%     \]
%     (Here $K_a$ is the per-wallet order count)
%     \item[iii.] \textbf{Non-owner wallet(s) execution.}
%     Either the owner does not execute the impactful sells,
%     \[
%     \forall a\in\mathcal{A}:\; a\neq O,
%     \]
%     \emph{or}, if the owner participates, every owner-originated sell stays sub-threshold:
%     \[
%     \max_{\,t_i:\,a_i=O}\; \mathrm{Impact}(t_i) \;\le\; \theta.
%     \]
%     (in the definition using in this research, we set the $\theta = 0.9$ as explained)
% \end{packeditemize}}

\begin{center}
    \fbox{
    \begin{minipage}{0.9\linewidth}
    \begin{packeditemize}
    \item[\ding{172}] \textbf{Control retention:} $\mathrm{burnLock}(\mathrm{LPT}_{\mathrm{init}})=\text{false}$. 
    The deployer maintains full control over the initial LP tokens.
    \item[\ding{173}] \textbf{Fragmented inflated selling.} 
    There exists a set of inflated seller(s) address $\mathcal{A}=\{a_1,\dots,a_N\}$ and, for each $a\in\mathcal{A}$, a sequence of sell events 
    $S_a=\{(t_{a,1},v_{a,1}),\dots,(t_{a,K_a},v_{a,K_a})\}$ such that
    \[
    \forall a\in\mathcal{A},\;\forall j\!\in\![K_a]:\quad \mathrm{Impact}(t_{a,j})=\frac{v_{a,j}}{\mathcal{V}(t_{a,j}^-)} \le \theta,
    \]
    (Here $K_a$ is the per-wallet order count)
    \item[\ding{174}] \textbf{Non-owner wallet(s) execution.}
    Either the owner does not execute the impactful sells,
    \[
    \forall a\in\mathcal{A}:\; a\neq O,
    \]
    \emph{or}, if the owner participates, every owner-originated sell stays sub-threshold:
    \[
    \max_{\,t_i:\,a_i=O}\; \mathrm{Impact}(t_i) \;\le\; \theta.
    \]
    (in the definition using in this research, we set the $\theta = 0.9$ as explained)
   \end{packeditemize}
    \end{minipage}
    }
\end{center}

Using the above actions, we label a pool as FRP when the deployer (i) retains LP-token control, while the extraction proceeds through either (ii) fragmented micro-trades or (iii) delegated wallets, or both. The pool thus avoids detection by canonical rug-pull heuristics, which rely on single large-impact or owner-originated exits, yet achieves the same economic outcome of draining the pool’s base-token reserves over time.

\end{defi}

\subsection{Threat Model}
\label{subsec:threat-model}

We focus on detecting and characterizing FRPs using only public on-chain data. In FRP games, two entities exist: the \textit{adversary} who performs the FRP, and \textit{defender}, represented by our measurement and detection pipeline.

\smallskip
\noindent\textbf{Adversary.}
An adversary controls one or multiple externally owned accounts and may also be the LP creator. Their capabilities reflect real-world observations: they can (i) create and seed liquidity pools, (ii) perform arbitrarily many inflated-sell transactions, and (iii) distribute sells across multiple wallets, and (iv) choose flexible timing and sell-volume patterns.  All operations follow standard AMM semantics and be recorded on-chain. The adversary cannot alter past state, forge events, or hide sender addresses and timestamps.

The adversary aims to drain liquidity while avoiding detection. To do so, they fragment exit volume, parallelize execution across wallets, and modulate time spans.

\smallskip
\noindent\textbf{Defender (our method).}
The defender represents our large-scale detection pipeline, which observes only public on-chain data: LP creation, liquidity changes, swap traces, seller wallets, and timestamps. No private keys, off-chain metadata, or centralized exchange information are assumed. 

The defender’s role is to identify FRP-like
patterns by analyzing (i) the set of participating wallets, (ii) the temporal
structure and magnitude of inflated sells, and (iii) the distribution of LPs across wallet-count and sell-count bins, as formalized in Algorithm~\ref{pseudo:frp-measurement}.

\smallskip
\noindent\textbf{Detection scope.} We restrict our model to AMM-based liquidity exit behaviors. Contract exploits, phishing, off-chain coordination, or oracle manipulation are outside scope. Our analysis captures only FRP behaviors that are fully observable through on-chain traces.

\begin{takeaway}
\noindent\textbf{Finding-(\S\ref{sec:def}):}
     Aggregating the evasive predicates yields a unified operational model of FRP. The model captures all practical variations as combinations of three execution parameters: number of seller wallets ($N$), per-wallet order count ($K_a$), and per-order volume ($v$) with optional owner participation. The only invariant needed to evade detection is the absence of owner transaction with $Impact > 0.9$.
\end{takeaway}

\begin{figure*}[h]
    \centering
    \includegraphics[width=0.95\linewidth]{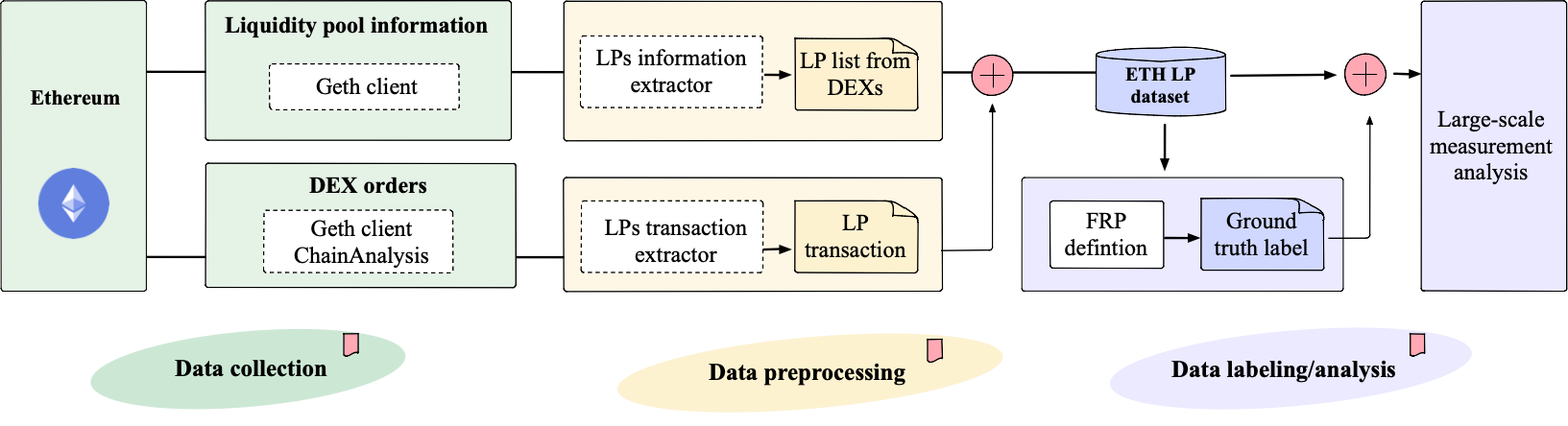}
    \caption{Overview of the research's large-scale measurements process of FRP LPs on Ethereum}
    \label{fig:process}
\end{figure*}

%\TODO{later add note on why we only use 100d life-length history}

\section{Large-Scale Measurement} 
\label{sec:large-scale}

We perform a large-scale empirical analysis of real-world FRP activities (visualized in Figure~\ref{fig:process}). Firstly, we use geth client and on-chain transaction API for LP data collection and preprocessing (\S\ref{subsec:data-prep}). Next, we apply the LP dataset with FRP definition (\S\ref{sec:def}) to systematically label liquidity pools that satisfy the evasive predicates of impact fragmentation and identity delegation (\S\ref{subsec:data-prep}). Finally, we conduct large-scale analysis on the flagged LPs (\S\ref{subsec:analyze_general},\S\ref{sec:results}).  

\subsection{Data Preparation}
\label{subsec:data-prep}

We construct a new dataset of confirmed scam pools, where each entry corresponds to one $\mathcal{P}$ and contains its complete transaction history from deployment to the final observed transaction.

\smallskip
\noindent\textbf{Data sources.}
We collect on-chain transactions from the six largest DEXs on Ethereum: Uniswap \cite{uniswap0}, Balancer \cite{balancer0}, PancakeSwap \cite{pancake0}, Curve Finance \cite{crv0}, BancorSwap \cite{bancor0} and SushiSwap \cite{sushi0} using the Geth client node \cite{geth}. 

We first collect all Ethereum transactions from block 6,627,000 
(Nov. 2, 2018, the date of the first LP deployment) to block 21,379,910 
(Mar. 5, 2025) using Geth client node \cite{geth}, and match them to LP deployment information using Chainalysis API~\cite{transpose}.  This yields a total of \textbf{384,029} liquidity pools across six major DEXes with at least one recorded activity within this range.

\smallskip
\noindent\textbf{Pool filtering.}
For these 384{,}029 pools, we further retrieve raw transaction data from a Geth archive node~\cite{geth} and categorize all interactions into DEX-related activities (\textit{deploy}, \textit{buy}, \textit{sell}, \textit{deposit}, \textit{withdraw}). 
Following our definition in \S\ref{sec:def}, we retain only pools whose observable lifetime 
(from first to last activity) does not exceed 100 days (details in \S\ref{subsec:tradeoff}). After applying this filter, \textbf{303,614} LPs remain.

\smallskip
\noindent\textbf{LP labeling.} 
We label each pool as an FRP instance based on the defined criteria. Applying this definition to the remaining 303,614 LPs, we identify \textbf{105,434} pools that match the patterns specified by our heuristic.

\begin{table}[b]
    \centering
    \renewcommand{\arraystretch}{1}
    \caption{Collected data on Ethereum.}
    \label{tab:datasummary}
    \begin{threeparttable}
        \resizebox{0.95\linewidth}{!}{
        \begin{tabular}{c|c|c|c}
            \toprule
            \multicolumn{2}{c}{\textbf{LPs}} & \multicolumn{2}{c}{\textbf{DEX Activities}}\\
            \cmidrule(lr){1-2}\cmidrule(lr){3-4}
            \multicolumn{1}{c}{DEXs} &
            \multicolumn{1}{c}{No. of pools} &
            \multicolumn{1}{c}{Order type} &
            \multicolumn{1}{c}{No. of Txs} \\
            \midrule
            Uniswap      & 376,854 &   &   \\
            SushiSwap    &   3,016 &  Deposit &  13,990,620 \\
            Balancer     &   2,932 &  Withdraw &  6,514,041 \\
            Curve        &     341 &  Buy & 588,462,142  \\
            PancakeSwap  &     462 &  Sell &  493,189,487 \\
            BancorSwap   &      424   &   &   \\
            \midrule
            \textbf{Total} & \textbf{384,029} &  & \textbf{1,102,156,290}\\
            \bottomrule
        \end{tabular}}
        \begin{tablenotes}
        \small \item[-] Abbreviation: \textbf{Tx}: transaction; \textbf{No.}: number.
        \end{tablenotes}
    \end{threeparttable}  
\end{table}

\smallskip
\noindent \textbf{Reproducibility.} To support research transparency and contribute for future research in experiment recreation, we release all scripts, analysis notebooks, and all code used in this study~\cite{frpsource}.

\begin{algorithm}
\caption{Large-Scale Measurement on Labeled FRP Pools}
\label{pseudo:frp-measurement}
\small
\SetAlgoLined

\KwData{\texttt{LPInfoData}, \texttt{LPTransactionData}}

\texttt{FRP\_LPs} $\gets$ \texttt{LPInfoData.label == 'FRP'}  
\label{line:select-frp} 
\\   \textcolor{olive}{/* select all LPs labeled as FRP */}

\BlankLine
Actor\_Centric\_Analysis(\texttt{FRP\_LPs}, \texttt{LPTransactionData}) 
\label{line:actor-call}
\\
Action\_Centric\_Analysis(\texttt{FRP\_LPs}, \texttt{LPTransactionData}) 
\label{line:action-call}
\\
Behavior\_Categorization(\texttt{FRP\_LPs}, \texttt{LPTransactionData}) 
\label{line:beh-call}
\\

%======================
%  ACTOR MODULE
%======================
\BlankLine
\SetKwProg{Fn}{\textcolor{blue}{Actor\_Centric\_Analysis} (\texttt{FRP\_LPs}, \texttt{LPTransactionData})}{:}{}
\Fn{}{
\label{line:actor-start}

    \ForEach{\texttt{P} in \texttt{FRP\_LPs}}{
        $S_P \gets$ set of wallets with at least one inflated sell in \texttt{P} \label{line:actor-walletset} \\
        $ownerInvolved_P \gets (\text{creator(P)} \in S_P)$  \label{line:actor-ownercheck}\\
        record $|S_P|$ and $ownerInvolved_P$ \label{line:actor-record}\\
    }
    compute distribution of $|S_P|$ over all LPs \label{line:actor-dist}\\
    compute yearly trend of single- vs multi-wallet exits \label{line:actor-yearly1}\\
    compute yearly fraction of owner-involved LPs \label{line:actor-yearly2}\\

\label{line:actor-end}
}

%======================
%  ACTION MODULE
%======================
\BlankLine
\SetKwProg{Fn}{\textcolor{blue}{Action\_Centric\_Analysis} (\texttt{FRP\_LPs}, \texttt{LPTransactionData})}{:}{}
\Fn{}{
\label{line:action-start}

    \ForEach{\texttt{P} in \texttt{FRP\_LPs}}{
        $T_P \gets$ ordered inflated-sell txs in \texttt{P} \label{line:action-txset}\\
        $\Delta t^{(P)}_{\text{first}} \gets$ time(first sell) $-$ time(LP creation) \label{line:action-dt1}\\
        $\Delta t^{(P)}_{\text{span}} \gets$ time(last sell) $-$ time(first sell) \label{line:action-dt2}\\
        $n^{(P)}_{\text{sell}} \gets |T_P|$ \label{line:action-sellcount}\\
        stratify metrics by cohorts: \{single/multi-wallet, owner/non-owner\} \label{line:action-stratify}\\
    }
    compute statistics used in Table~\ref{tab:cohort-board} \label{line:action-stats}\\

\label{line:action-end}
}

%======================
%  BEHAVIOR MODULE
%======================
\BlankLine
\SetKwProg{Fn}{\textcolor{blue}{Behavior\_Categorization} (\texttt{FRP\_LPs}, \texttt{LPTransactionData})}{:}{}
\Fn{}{
\label{line:beh-start}

    \ForEach{\texttt{P} in \texttt{FRP\_LPs}}{
        $w \gets$ number of seller wallets in \texttt{P} \label{line:beh-w}\\
        $c \gets$ total inflated-sell count in \texttt{P} \label{line:beh-c}\\
        assign \texttt{P} to wallet-bin$(w)$ and sell-bin$(c)$ \label{line:beh-bin}\\
    }
    compute share of LPs in each (wallet-bin, sell-bin) \label{line:beh-share}\\
    label dominant behavior regions as:  
    \{Minimal Drains, Moderate Networks, Distributed Campaigns\} \label{line:beh-label}\\
    generate category summaries (Fig.~\ref{fig:wallet-sell-scatter}, 
            \ref{fig:sell_dist}, \ref{fig:seller_dist}, 
            Table~\ref{tab:inflated-category}) \label{line:beh-summary}\\

\label{line:beh-end}
}

\end{algorithm}

\subsection{Dataset Analysis}\label{subsec:analyze_general}

We provide a brief review of our labelled FRP dataset.

The dataset contains \textbf{105,434} LPs deployed between 2018 and 2024, with the number of detected cases rising sharply after mid-2020. Figure~\ref{fig:label_lps} presents the yearly distribution in detail. FRP activity is negligible in 2018–2019 (only 4 and 76 LPs), but increases by an order of magnitude with the expansion of DeFi after mid-2020: 8,195 LPs in 2020 and 8,524 in 2021. The growth accelerates further in 2022 (21,455) and peaks in 2023 (39,351). In Figure~\ref{fig:label_lps}, the LP counts stop in 2024 (27,829) without 2025 data due to the partial observation window since our transaction (data collected up to March 2025, which later-formed LPs may not yet exhibit full exit behavior and lead to only validated FRP with the latest created time of Dec 2024) and an overall market slowdown~\cite{slowdown}. Overall, over 90\% of labelled LPs appear after 2020, indicating that FRP is a recent phenomenon concentrated in 2022–2024.

Across all labeled pools, we record \textbf{1,501,408} unique users and \textbf{34,192,767} transactions. On average, each pool interacts with 14.24 users and involves 324.30 transactions. A total of 401,838 wallets are identified as inflated sellers, representing 26.76\% of all interacting addresses, meaning more than one in four participants engage directly in manipulative behavior. The average pool lifespan is 14.47 days. The underlying distributions of those averages are highly skewed: most pools are much smaller than the mean, while a small number of large operations dominate total activity.

\subsection{Implementations}
\label{subsec:appendixpseudo}

To make our measurement pipeline transparent, we summarize the end-to-end workflow in Algorithm~\ref{pseudo:frp-measurement}. 

Lines~\ref{line:select-frp}–\ref{line:actor-end} extract the set of LPs labeled as FRP and compute wallet-level properties such as the number of seller wallets and whether the pool creator participates in inflated sells. Lines~\ref{line:action-start}–\ref{line:action-end} encode temporal and structural metrics of the inflated-sell trace, enabling stratified comparisons across single-wallet vs. multi-wallet and owner- vs. non-owner cohorts.  Lines~\ref{line:beh-start}–\ref{line:beh-end} map each pool into wallet- and sell-based bins, from which we derive the three behavioral clusters (Minimal Drains, Moderate Networks, Distributed Campaigns) reported in Figures~\ref{fig:wallet-sell-scatter}–\ref{fig:seller_dist} and Table~\ref{tab:inflated-category}.

\section{Evaluation and Analyses} 
\label{sec:results}

The labelled pools are examined from two complementary perspectives: 

\begin{packeditemize}
    \item \textit{Actor-centric analysis} (\S\ref{subsec:actor_centric}) investigates \emph{who} performs the draining actions, quantifying wallet distribution, ownership participation, and their evolution over time.  
    \item \textit{Action-centric analysis} (\S\ref{subsec:action-centric}) explores \emph{how} draining is executed by analyzing sell frequency, fragmentation levels, and behavioral categories.  
\end{packeditemize}

We then integrate these perspectives to reveal the aggregate scale and temporal dynamics of FRPs (\S\ref{subsec:inflated-categories}).

\subsection{Actor-Centric Analysis}
\label{subsec:actor_centric}

This analysis examines the actors behind FRP scam, specifically, \textit{how many wallets participate in each drain}, \textit{what roles they play}, and \textit{how profits are distributed among them}.
By characterizing who performs the drain, we aim to understand the attacker's preferences, whether they in general prefer relying on a single dominant wallet or intentionally spread activity across many smaller addresses to evade heuristic detection.

For each labelled pool, we collect all addresses that perform at least one inflated sell order. We then compare these addresses with the deployer (owner) wallet and classify the remaining sellers as delegated or external wallets. Finally, we analyze the distribution of wallet counts across all pools.

\smallskip
\noindent\textbf{Overall distribution.} Across the full corpus, single-seller drains remain the most common pattern when pools that got inflated sell by only 1 seller account for 28.04\% of all labeled LPs (29,573 LPs). It is important to note that these single-seller cases do not correspond to canonical rug-pulls: by construction, all known rug-pull instances and high-impact owner sells were excluded during data preparation by applying FRP definition. Here, a “single-seller” FRP simply denotes scams in which only one wallet executes all inflated sells, yet the transaction(s) either remain below the impact threshold (Predicate~\eqref{eq:B}) or originate from a non-owner wallet (Predicate~\eqref{eq:C}).

Most FRP schemes still operate through small teams: pools with inflated seller counts between 1 and 10 account for 70.45\% of all cases, totaling 74{,}287 LPs. The remaining 29.55\% (31{,}147 LPs) form a long tail of pools with more than ten sellers, suggesting coordinated campaigns that deliberately distribute execution across larger address sets.

% \begin{figure}[h]
%   \begin{minipage}[t]{\linewidth}
%     \centering
%     \includegraphics[width=\linewidth]{graphs/distribution_change.png}
%     \caption{Shifting in involved wallets that participate in inflated selling among labeled LPs}
%     \label{fig:distribution_change}
%   \end{minipage}
%   \hfill
%   \begin{minipage}[t]{\linewidth}  
%     \centering
%     \includegraphics[width=\linewidth]{graphs/owner_distribute.png} 
%     \caption{Proportion of LPs with owner address exits}
%     \label{fig:owner_distribute}
%   \end{minipage}
% \end{figure}

% \begin{figure}[!]
% \centering
% \resizebox{\linewidth}{!}{
% \begin{tikzpicture}
% \begin{axis}[
%   width=13cm,height=5cm,
%   ybar, bar width=0.35cm,
%   symbolic x coords={2017,2018,2019,2020,2021,2022,2023,2024},
%   xtick=data,
%   xticklabel style={rotate=30, anchor=east, yshift=-2pt},
%   xlabel={Year},
%   ylabel={FRP LPs ($\times 10^{4}$)},
%   ymin=0, ymax=50000,
%   scaled y ticks=false,
%   ytick={0,10000,20000,30000,40000,50000},
%   yticklabels={0,1,2,3,4,5},
%   nodes near coords,
%   every node near coord/.append style={yshift=5pt},
%   nodes near coords style={/pgf/number format/.cd, fixed, precision=0, 1000 sep={,}},
%   enlarge x limits=0.2,
%   grid=both, grid style={dotted},
% ]

% \addplot[pattern=north east lines, pattern color=blue] coordinates {
%     (2018, 4)
%     (2019, 76)
%     (2020, 8195)
%     (2021, 8524)
%     (2022, 21455)
%     (2023, 39351)
%     (2024, 27829)
% };
% \end{axis}
% \end{tikzpicture}
% }
% \caption{Statistic of the created FRP LPs over years.}
% \label{tab:label_lps}
% \end{figure}

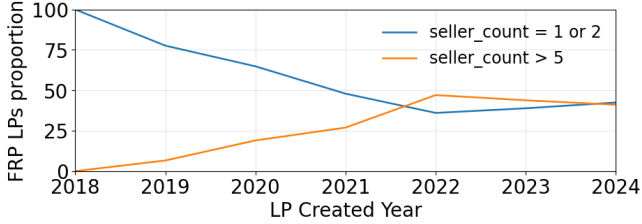
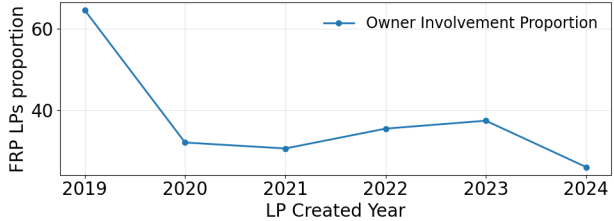
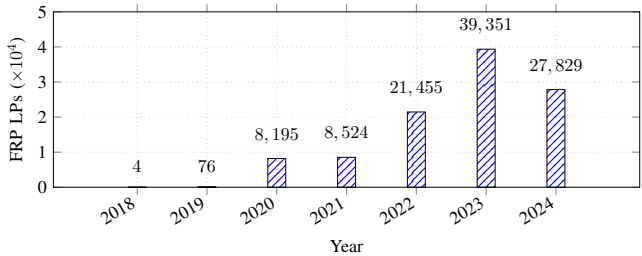
\begin{figure}[t]
  \subfigure[Shifting in involved wallets used to participate in inflated selling in each LPs among labeled FRP]{%
    \label{fig:distribution_change}
    \begin{minipage}[b]{\linewidth}
      \centering
      \includegraphics[width=1\linewidth]{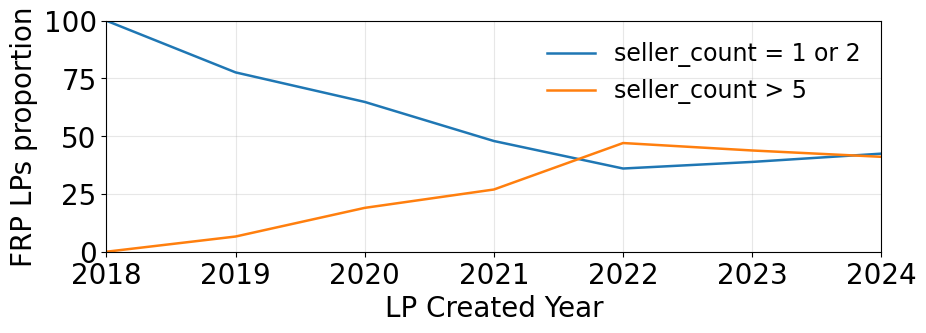}
    \end{minipage}
  }

  \vspace{5pt}

  \subfigure[Proportion of FRP LPs with owner address participated in inflated selling exits]{%
    \label{fig:owner_distribute}
    \begin{minipage}[b]{\linewidth}
      \centering
      \includegraphics[width=0.95\linewidth]{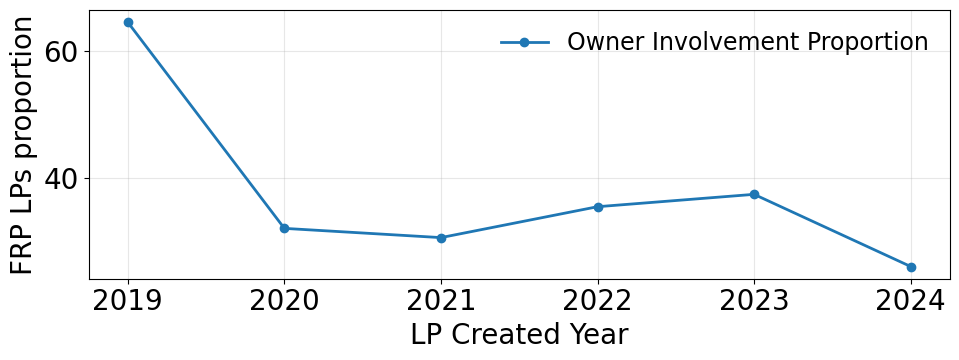}
    \end{minipage}
  }

  \vspace{5pt}

  % (c) yearly FRP counts (TikZ)
  \subfigure[Number of FRP-labeled LPs created by year]{%
    \label{fig:label_lps}
    \begin{minipage}[b]{\linewidth}
      \centering
      \resizebox{\linewidth}{!}{%
      \begin{tikzpicture}
        \begin{axis}[
          width=13cm,height=5cm,
          ybar, bar width=0.35cm,
          symbolic x coords={2017,2018,2019,2020,2021,2022,2023,2024},
          xtick=data,
          xticklabel style={rotate=30, anchor=east, yshift=-2pt},
          xlabel={Year},
          ylabel={FRP LPs ($\times 10^{4}$)},
          ymin=0, ymax=50000,
          scaled y ticks=false,
          ytick={0,10000,20000,30000,40000,50000},
          yticklabels={0,1,2,3,4,5},
          nodes near coords,
          every node near coord/.append style={yshift=5pt},
          nodes near coords style={/pgf/number format/.cd, fixed, precision=0, 1000 sep={,}},
          enlarge x limits=0.2,
          grid=both, grid style={dotted},
        ]
          \addplot[pattern=north east lines, pattern color=blue] coordinates {
            (2018, 4)
            (2019, 76)
            (2020, 8195)
            (2021, 8524)
            (2022, 21455)
            (2023, 39351)
            (2024, 27829)
          };
        \end{axis}
      \end{tikzpicture}
      }% end resizebox
    \end{minipage}
  }

  \caption{Temporal and actor-side evolution of FRP scams}
  \label{fig:actor_triple}
\end{figure}

\smallskip
\noindent\textbf{Transition from single- to multi-wallet exits.} When looking at the yearly distribution of inflated seller count among the pool by created day, we can observe that the distribution has shifted materially over time, which indicates the changing of preferred fraud scheme toward multiple-wallet drains. 
In 2019, single-seller drains accounted for 57\% of all cases, but their share declined steadily, reaching 28.3\% by the end of 2024.
Figure~\ref{fig:distribution_change} illustrates this trend in detail: between 2018 and mid-2021, scams involving multiple selling wallets began to surpass traditional single-wallet exits. After 2021, the proportions of small-wallet (1–10 sellers) and large-wallet (10+ sellers) drains stabilized at roughly 35–45\%, indicating that both strategies now coexist.
Some attackers continue to rely on a single controlled wallet for operational simplicity, while others increasingly adopt multi-wallet coordination to evade heuristic detection.
Overall, the data confirm a clear shift from single-wallet to multi-wallet execution, consistent with the identity-evasion strategy discussed in \S\ref{subsec:modify-c}.

\smallskip
\noindent\textbf{Owner wallet involvement.} Among the flagged LPs, 33.13\% (34,927 LPs) recorded the involvement of LP's owner address in committing inflated selling activity. Figure~\ref{fig:owner_distribute} illustrates the temporal trend of owner wallet participation among flagged pools. The data show a sharp decline in direct owner involvement after 2019, dropping from over 65\% of pools in early datasets to roughly 25-30\% in 2020-2022, and stabilizing near this lower range through 2023. A minor rebound is visible in 2022–2023, suggesting that some attackers intermittently returned to partial owner participation, likely using low-impact or mixed-owner sells that remain below detection thresholds. By 2024, however, the share falls again to its lowest point (24\%), confirming that most recent FRP scams now shifting toward the trend of avoiding the involvement deployer wallet in their exit. 

This pattern aligns with the broader shift toward multi-wallet and delegated execution discussed earlier: as public awareness and heuristic tools increasingly flag large owner-originated sells, scammers appear to have systematically phased out direct owner involvement.

\smallskip
\noindent\textbf{Wallets that are involved in multiple LPs.} We further examine wallets that appear in more than one flagged LP, i.e., any address that conducts inflated selling in at least two pools. Among the 401,838 distinct wallets involved in inflated selling, 127,252 (31.67\%) are recurrent, while 274,586 (68.33\%) occur only once. Table~\ref{tab:top-recurrent-wallets} lists the five most active recurrent wallets and their LP participation counts. Recurrence is highly concentrated. 

The most active address (0xae2...FaE13) appears in 16,525 flagged LPs, more than twice the count of the next most active wallet (0x77a...C65d1, with 8,097 LPs). All top-10 wallets exceed 2.4k appearances, the top-20 remain above 1.4k, and even the 50th most active wallet still shows 662 occurrences. However, recurrence does not necessarily correlate with illicit profit: several wallets involved in fewer LPs earned higher gains (see Table~\ref{tab:top-recurrent-wallets}).

\begin{table}[http]
    \centering
    \renewcommand{\arraystretch}{1}
    \caption{Top Addresses by FRP Pool Participation}
    \label{tab:top-recurrent-wallets}
    \resizebox{\linewidth}{!}{
    \begin{tabular}{>{\centering\arraybackslash}p{3.5cm} |>{\centering\arraybackslash}p{3.5cm}  c}
        \toprule
        \multicolumn{1}{c}{\makecell{\textbf{Scammer}\\{\textbf{address}}}} & \makecell{\textbf{FRP inflated}\\{\textbf{selling involved}}} & \makecell{\textbf{Profit}\\{\textbf{(USD)}}}  \\
        \midrule         
        0xae2...FaE13 & 16525 & \$9,009,675\\
        0x77a...C65d1& 8097 &\$1,276,803\\
        0x3fC...b7FAD& 3303 &\$79,378,720\\
        0xE59...61564& 2798 &\$59,945,861\\
        0xf9c...ABE68& 2736 &\$466,592\\
        \bottomrule
    \end{tabular}
    }
\end{table}

\subsection{Action-Centric Analysis}
\label{subsec:action-centric}

We analyze the behavioral dynamics of inflated selling within flagged liquidity pools. We contrast two operational axes: (i) \textit{single-} v.s. \textit{multi-wallet} execution, which captures whether the attack is conducted through a single dominant address or multiple coordinated ones, and (ii) \textit{owner} v.s. \textit{non-owner} participation, reflecting whether the deployer wallet itself engages in the manipulation. Each combination thus represents a behavioral cohort of strategies.

\vspace{0.5em}
\noindent\textbf{Measurement methodology.}
For each LP $p$, we extract three temporal and activity-based features: 
\begin{packeditemize}
    \item The \emph{first-sell delay} $\Delta t^{(p)}_{\text{first}} = \min(t^{(p)}_{\text{sell}}) - t^{(p)}_{\text{create}}$, measuring the onset of draining relative to pool creation.
    \item The \emph{sell span} $\Delta t^{(p)}_{\text{span}} = \max(t^{(p)}_{\text{sell}}) - \min(t^{(p)}_{\text{sell}})$, quantifying the duration of sustained selling.
    \item The \emph{number of inflated sells} $n^{(p)}_{\text{sell}}$, representing the total intensity of selling time. 
\end{packeditemize}

These measurements reveal how aggressively and persistently each LP is exploited. We show our results in Table~\ref{tab:cohort-board} and explain them as below.

\begin{table}[h]
  \centering
  \caption{Selling Behavior across Wallet and Ownership}
  \label{tab:cohort-board}
  \resizebox{\linewidth}{!}{
  \setlength{\tabcolsep}{6pt}
  \begin{tabular}{c|cccc}
    \toprule
    & \multicolumn{2}{c}{Single Wallet} & \multicolumn{2}{c}{Multiple Wallets} \\
    \cmidrule(lr){2-3}\cmidrule(lr){4-5}
    Metric &
      Owner & Non-owner &
      Owner & Non-owner \\
    \midrule
    No. LPs                  & 11,613 & 17,960 & 23,314 & 52,547 \\
    Avg First-sell (days)        & 2.16 & 1.55 & 0.68 & 0.84  \\
    Avg No. Sell        & 3.34 & 3.38 & 82.47 & 96.62 \\
    Avg Sell Span (days)        & 0.45 & 0.24 & 11.76 & 6.34 \\
    \bottomrule
  \end{tabular}
  }
\end{table}

\noindent\textbf{Behavioral trends.}
Single-wallet scams are characterized by rapid and compact liquidation phases. The median first-sell occurs within approximately 1.5–2 days after deployment, and selling activity typically concludes within a narrow window (median span below one day). The strategy is largely scripted and executed shortly after liquidity provisioning, with minimal attempt to disguise exit behavior. The average number of inflated sells remains low (around 3.4 transactions per pool), consistent with a single address completing most of the drain in one or two bursts of activity.

In contrast, multi-wallet LPs exhibit a different temporal pattern. Their first-sell actions occur sooner, often within the first 12–20 hours on average, but their selling extends for significantly longer, with mean spans stretching over 6–12 days. This elongated pattern indicates that multi-wallet setups distribute draining activity over time, potentially to evade transaction-level heuristics or simulate organic user interaction. The total number of sell transactions per pool rises dramatically (by more than an order of magnitude), suggesting a clear operational distinction: single-wallet scams are opportunistic and direct, whereas multi-wallet scams are structured, strategic, and persistent.

\vspace{0.5em}
\noindent\textbf{Owner v.s. non-owner participation.}
Comparing owner-involved and non-owner LPs reveals systematic timing differences. Pools where the deployer directly participates in selling tend to initiate liquidation earlier, reflecting the attacker’s immediate control and awareness of deployment timing. In contrast, non-owner executions exhibit longer sell spans, consistent with delegated draining that unfolds gradually over time. This pattern suggests an obfuscation strategy, in which the deployer’s address abstains from direct selling to reduce traceability, transferring exit operations to auxiliary wallets instead.

\begin{figure}[t]
    \subfigure[The number of total distinct seller wallets in all flagged FRP LPs]{\label{fig:seller_dist}
    \begin{minipage}[b]{\linewidth}
    \centering
    \includegraphics[width=1\linewidth]{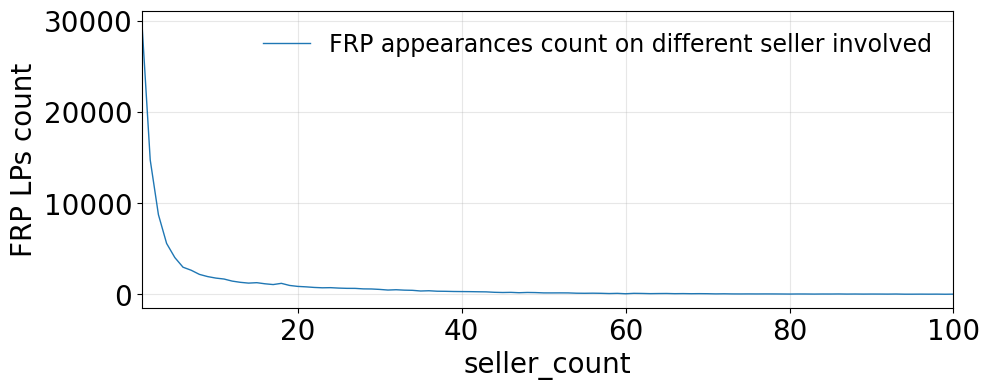}
    \end{minipage}
    }
    \vspace{5pt}
    \subfigure[The number of distinct inflated sell order among all flagged FRP LPs]{ \label{fig:sell_dist}
    \begin{minipage}[b]{\linewidth}
    \centering
    \includegraphics[width=1\linewidth]{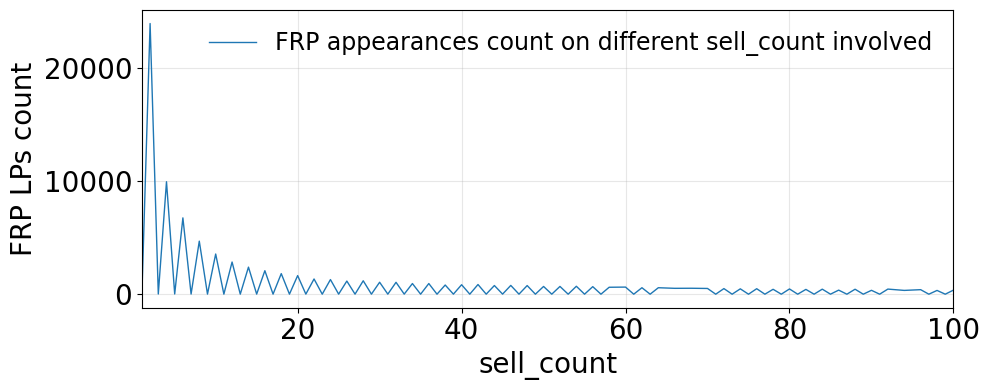}
    \end{minipage}
    }
    
    \caption{Distribution of inflated selling wallets and inflated sell counts in all of flagged FRP LPs}
    \label{fig-howtransfer}
\end{figure}

\subsection{Categorization of FRP Behaviors}
\label{subsec:inflated-categories}

We extend our analysis to structural characteristics observable across all LPs, covering both single- and multi-wallet LPs. Each pool is characterized by its number of unique wallets conducting inflated selling ($\mathsf{seller\_count}$) and the total number of sell transactions ($\mathsf{sell\_count}$). 

Figure~\ref{fig:seller_dist} shows the frequency distribution of the number of distinct seller wallets ($\mathsf{seller\_count}$) involved in each pool (from 1 to 100 sellers for demonstration). 
The distribution is highly right-skewed, which indicates that while many scammers still adopt minimal wallet setups, a growing subset deliberately fragments their selling activity across a large network of addresses, likely to avoid identity-based detection (Predicate~\eqref{eq:C}).

Figure~\ref{fig:sell_dist} further plots the number of sell orders ($\mathsf{sell\_count}$) executed per pool (from 1 to 100 inflated sell for demonstration). Similar to the seller distribution, most scams exhibit only a handful of selling transactions, yet a small fraction perform extensive sequences of sales (sometimes exceeding fifty per pool) demonstrating systematic fragmentation (Predicate~\eqref{eq:B}).

Together, these distributions highlight the heterogeneity of execution patterns among evasive scams, motivating the actor- and action-centric analyses that follow. In addition, to capture more meaningful behavioral space, we bin both axes into interpretable ranges:
\begin{itemize}
    \item \textit{Wallet bins:} 1, 2–4, 5–9, and 10+;
    \item \textit{Sell bins:} 0–9, 10–49, 50–249, and 250+.
\end{itemize}

\begin{figure}[t]
    \centering
    \includegraphics[width=0.95\linewidth]{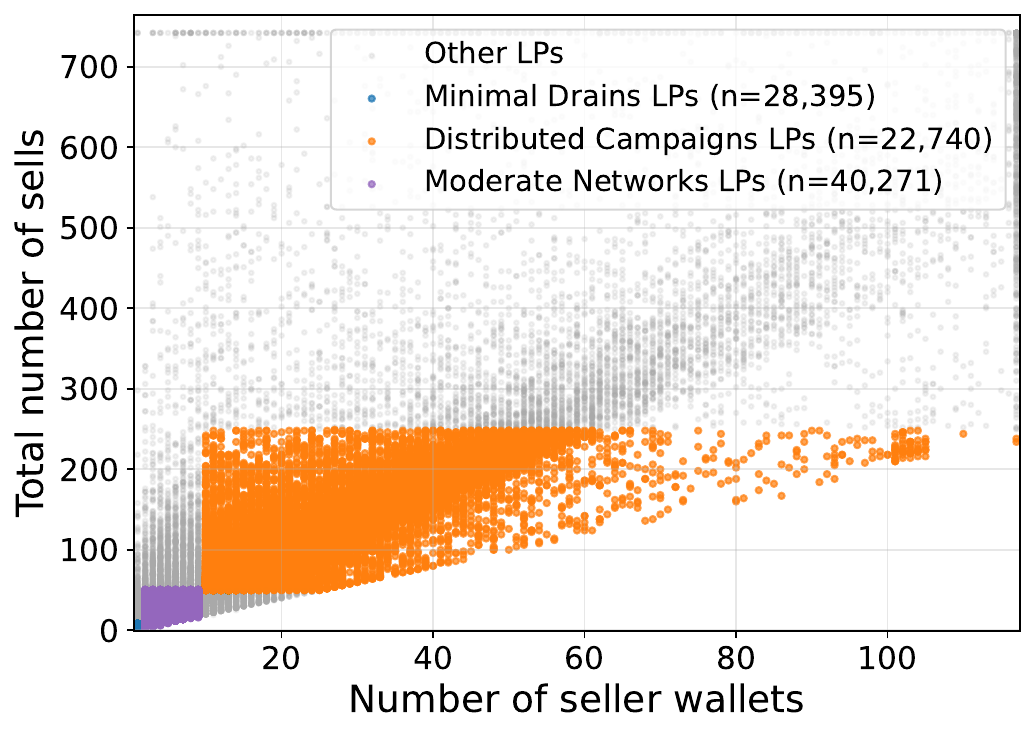}
    \caption{Scatter distribution of scam LP behaviors by wallet count and sell activity.}
    \label{fig:wallet-sell-scatter}
\end{figure}

The resulting 4$\times$4 grid summarizes LP behaviors, with each bin’s share representing its percentage among all detected scam pools.
Figure~\ref{fig:wallet-sell-scatter} visualizes this distribution, where each point corresponds to a flagged LP positioned by its number of seller wallets ($x$-axis) and total sell transactions ($y$-axis).
Points are color-coded to highlight behavioral categories:
\textcolor{blue!70!black}{Minimal Drains},
\textcolor{orange!80!black}{Distributed Campaigns}, and
\textcolor{purple!70!black}{Moderate Networks}.
Residual pools outside these clusters are shown in light gray.
Table~\ref{tab:inflated-category} further summarizes the dominant regions within this behavioral space.

Three dominant behavioral categories emerge, together accounting for over 75\% of all identified scam LPs:

\begin{packeditemize}
    \item \textit{Minimal Drains (26.9\%)}. Pools with a single wallet and fewer than ten sell transactions.  
    These represent the simplest form of exit scam: a one-operator event involving minimal coordination and a short time horizon. 
    The actor typically drains liquidity soon after deployment with little evidence of distribution. 

    \item \textit{Distributed campaigns (21.6\%)}. Pools operated by large coordinated groups ($\geq$ 10 wallets) with moderate sell activity (50–249 transactions).  
    These scams exhibit structured execution and likely prefer to rotate wallets or fragment sales, maintaining liquidity appearance before completing their LP extraction process. 
    Their scale and repetition suggest sustained, systematized operations.

    \item \textit{Moderate networks (27.3\%)}. Formed by mid-sized groups (2–9 wallets) engaging in up to 50 sells per pool.  
    These LPs combine mild coordination with measured selling activity, possibly reflecting adaptive behaviors of scammers scaling from individual to distributed operation. 
\end{packeditemize}

A smaller fraction ($<25\%$) of LPs lie outside, either exhibiting irregular patterns or incomplete trading records.

\begin{table}[t]
    \centering
    \caption{Dominant inflated-selling categories.}
    \label{tab:inflated-category}
    \resizebox{\linewidth}{!}{
    \begin{tabular}{c|cc}
        \toprule
       \multicolumn{1}{c}{\textbf{Category}} & \textbf{Definition (Wallets, Total sells)} & \textbf{Share of LPs} \\
        \midrule
        Minimal Drains & 1 wallet, 0–9 sells & 26.9\% \\
        Distributed Campaigns & $\geq$ 10 wallets, 50–249 sells & 21.6\% \\
        Moderate Networks & 2–9 wallets, $\leq$ 50 sells & 27.3\% \\
        \midrule
        \textbf{Other} & - & $\approx$24.1\% \\
        \bottomrule
    \end{tabular}
            }
\end{table}

%This structural typology highlights a consistent scaling pattern in inflated-selling behavior.  A noticeable number of scams cluster in low-effort, single-wallet operations, but a significant proportion remain and show evidence of multi-address coordination.  This transition from individual to distributed execution reflects the evolution of scam liquidity management: wallet rotation enables repeatable draining strategies at scale.

\begin{takeaway}
\noindent\textbf{Finding-(\S\ref{sec:large-scale}):}
 We identified 105,434 FRP pools and found that evasive exits are frequent.
\begin{packeditemize}
\item  Single-seller scams remain the majority but show a steady decline (from 57\% to 28.3\% between 2019 and 2024), while multi-wallet exits have risen continuously and now account for nearly half of all cases.
\item  Owner involvement has decreased from 65\% to 24\%, reflecting a growing preference for delegated execution through secondary wallets.
\item  Temporal and behavioral analyses reveal two patterns: (i) rapid, single-wallet drains; (ii) slow, distributed exits emphasizing stealth. 
\end{packeditemize}

\end{takeaway}
%  Applying our formal definition to 303,614 Ethereum LPs, we identify 105,434 fragmented rug pull (FRP) pools, confirming that evasive exits are both frequent and systematic. Single-seller scams remain dominant but declining (57\%→28.3\% from 2019–2024), while multi-wallet exits have steadily increased, now representing nearly half of all cases. Owner involvement dropped from 65\% to 24\%, indicating a growing preference for delegated execution. Temporal and behavioral analyses further reveal two stable archetypes: rapid, single-wallet drains for immediacy and slow, distributed exits for stealth. These findings empirically validate that the evasive modifications theorized in~\ref{sec:proving} and formalized in~\ref{sec:def} now represent the mainstream form of on-chain liquidity fraud.

\section{Discussion}
\label{sec:mitigation}

\subsection{Mitigation and Early-Warning}
Fragmented rug pull remain difficult to identify in \textit{real time} because they mimic legitimate trading dynamics and distribute execution across multiple addresses. We still recommend a series of proactive checks for mitigation.

\begin{packeditemize}
  \item \textbf{Verify token intent and behavioral consistency.}  
  Cross-check the token’s stated utility (e.g., governance, staking, yield) with its on-chain activity.  
  Inconsistencies between declared purpose and observed liquidity actions, such as continuous sell pressure or recurring distributor creation, should be treated as risk indicators.

  \item \textbf{Evaluate LP-token status.}  
  Confirm that the initial LP tokens are either burnt or locked under verifiable mechanisms.  
  Unburned transferable LP tokens remain the strongest structural prerequisite for any rug pull variant. Open-source tools like Etherscan, DEXTools, and Chainalysis can assist with verification.

  \item \textbf{Monitor gradual liquidity changes.}  
 FRP drain liquidity over hours or days rather than instantly. Track LP balances and ownership over time; repeated small sells coupled with a steady net liquidity decline are strong indicators of an ongoing stealth exit.

  \item \textbf{Leverage cross-wallet forensics.}  
  FRP rely on many distributor addresses. Apply clustering and linkage heuristics to detect shared funding sources, synchronized transaction timings, or repeated reuse of intermediary addresses. Integrate these relational signals into monitoring dashboards or analytic APIs for better early warning capability.
\end{packeditemize}

We note that these methods emphasize pattern-oriented detection rather than isolated event alerts. This broader, behavior-based view allows systems to capture fuzzy yet recurring anomalies, aligning with our observation that fragmented rug pulls evade per-transaction heuristics but become visible through aggregated behavioral trends.

\subsection{Comparison to Related Concepts}
\label{sec:related-phenomena}

\noindent\textbf{Stock dilution.} 
A fragmented rug pull shares conceptual parallels with unauthorized stock dilution in traditional finance, where issuers increase the circulating supply without shareholder approval or disclosure. Regulatory authorities such as the U.S. Securities and Exchange Commission (SEC) strictly prohibit unapproved share issuance and naked short selling~\cite{nysedefinition, morganstanleyequity}. For instance, SEC sanctioned Goldman Sachs for facilitating trades that created synthetic stock exposure without regulatory consent~\cite{nysedefinition}.

Similarly, in a fragmented rug pull, token deployers retain unilateral control over token supply and can gradually reduce existing holders’ value through incremental sell-offs. Although executed via decentralized smart contracts, this process follows the same principle as illicit stock issuance: expanding one’s redeemable share while bypassing consent and transparency requirements~\cite{seccase1,seccase2}.

\smallskip
\noindent\textbf{MEV bots.}
FRP may appear similar to automated arbitrage activity because both involve frequent, small-volume trades that gradually adjust liquidity pool balances. However, their mechanisms and intentions are fundamentally different.

MEV bots are autonomous agents that exploit short-lived price discrepancies across pools or exchanges~\cite{weintraub2022flash,li2023demystifying}. Their trades are economically neutral, driven by transparent, rule-based strategies that restore price equilibrium rather than extract value through deception. In contrast, a fragmented rug pull deliberately splits large sell orders and distributes them across multiple wallets to conceal coordinated liquidity drainage.
While arbitrage bots stabilize markets by realigning token ratios, FRP destabilize them by continuously removing real liquidity through inflated selling. Moreover, arbitrage agents have no privileged access to LP tokens or minting rights. Their profits arise from market inefficiencies, not control over supply.

\subsection{Applications To DeFi Communities}\label{subsec:application}

Beyond detection, our FRP model enables \emph{preventive} mechanisms for AMMs. Because FRP behavior must satisfy specific structural predicates, DEXes can introduce lightweight safeguards such as adaptive fees for abnormal micro-sell bursts or temporary throttling of suspicious multi-wallet activity. These require no off-chain signals and can operate transparently within permissionless pools.

Our behavioral taxonomy also informs ecosystem tools. Wallets and DEX aggregators can expose FRP-risk indicators to end users, while analytics platforms can incorporate our wallet–sell distributions to flag unstable pools. At a broader level, the observed cross-wallet coordination patterns support market-level scam modeling and regulatory investigations, offering a new lens for understanding liquidity-drain dynamics across DeFi networks.

\subsection{Design Trade-Off}\label{subsec:tradeoff}
Our measurement targets liquidity pools with observable lifetimes shorter than 100 days, as a trade-off between coverage and precision in identifying fraudulent pools.

Empirically, short-lived pools are highly symptomatic of malicious or fraudulent behavior. Large-scale analyses \cite{chainanalysis2023report}\cite{chainanalysis2024report}\cite{rugdocreport} show that most known rug pull or scam tokens collapse within weeks of deployment, with a median lifetime below 30 days. By contrast, legitimate DeFi projects sustain trading activities well beyond this period.

We therefore select 100 days as an operational boundary to balance two competing objectives:

\begin{packeditemize}
    \item reducing false negatives by capturing the majority of short-lived, high-risk pools; and
    \item avoiding false positives from legitimate tokens that exhibit natural liquidity fluctuations over longer horizons.
\end{packeditemize}

Although some evasive FRP variants may prolong their operation through slower or staged liquidity drains, expanding the observation window would admit numerous benign cases (e.g.,seasonal liquidity shifts or post-airdrop volatility) that do not indicate coordinated exits.

\section{Related Works on DeFi Fraud Detections}
\label{sec:related}

Earlier studies established broad taxonomies of DeFi scams~\cite{zhou2023sok}, We present representive ones.

\smallskip
\noindent\textbf{Rug pull and Honeypot.}
Rug pulls remain the most studied DEX-specific scam, where liquidity providers abruptly withdraw their tokens to drain a pool’s reserves.
Bruno~et~al.~\cite{rugpull} applied machine learning to identify rug-pull and scam-prone tokens based on transaction histories and contract metadata. Cernera~et~al.~\cite{spammer} conducted a longitudinal analysis of rug-pull lifetimes on Ethereum and BNB Smart Chain.

Honeypots embed malicious logic into the token’s smart contract, trapping buyers who attempt to sell back their tokens and thereby enabling the deployer to seize their funds.
Such contract-level frauds have drawn extensive attention from the research community~\cite{honeypot,honeypot2,honeypot3,honeypot4}.
Rundong~et~al.~\cite{honeypot2} proposed heuristic-based honeypot detection, and Christof~et~al.~\cite{honeypot} expanded symbolic-execution models to identify malicious logic before deployment.

However, these approaches rely on static rule-based heuristics, i.e., flagging a single large owner withdrawal or a high-impact sell. We move beyond static detection by introducing a dynamic as well as fine-grained framework that models both temporal and multi-actor behaviors.  

\smallskip
\noindent \textbf{More on-chain scams.}  
Beyond DEX-specific studies, broader research for on-chain DeFi crimes examined a wide spectrum of manipulative behaviors, including pump-and-dump events~\cite{pd1,pd2}, Ponzi and pyramid schemes~\cite{sadponzi,ponzi2}, and decentralized money-laundering patterns used by scammers to conceal illicit gains~\cite{lin2024denseflow}.  
NFT-related rug pulls~\cite{nftrug} also attracted attention due to the massive economic impact of the NFT ecosystem~\cite{wang2021non}.  

Another emerging category is the counterfeit token scam, where attackers issue tokens under the same name as legitimate cryptocurrencies to deceive investors~\cite{counterfeitscam}.  
In addition, several niche but profitable scams (e.g., frontrunning~\cite{frontrunningscam}, role-play frauds~\cite{roleplayscam}) have been investigated to decode their operational signatures.  
Malicious actors continuously modify transaction behaviors to evade static detection heuristics, and capturing new scams is always challenging.

\smallskip
\noindent \textbf{Scam detection.}  
Blockchain scam detection has evolved from early rule-based heuristics to more contextual and data-driven approaches. Traditional detectors rely on static signals (e.g., large owner sells, sudden liquidity withdrawals, or rapid price drops~\cite{rugpull,spammer,honeypot,honeypot2}) but their fixed thresholds and
dependence on known actor identities make them easy to evade and mostly useful only after the scam has unfolded. Recent work~\cite{slid} introduces temporal features and adaptive logic to better capture  manipulation. Parallel efforts use graph and neural models to identify abnormal wallet behavior or
collusive structures~\cite{graph1,graph2,graph3,graph4}, though these methods require large labelled datasets and often lack interpretability when applied to LP dynamics.

\smallskip
\noindent \textbf{Ours!}  To our knowledge, no prior study formally decomposes rug-pull heuristics into underlying atomic conditions.  We conceptually extend their scope and identify a rich set of evasive strategies that have been previously overlooked.
%Existing studies provide extensive scam taxonomies but offers limited formalization of how scammers adapt their behaviors to evade detection.  
\section{Conclusion}

This paper introduces the \textit{fragmented rug pull} (FRP) scam, a family of evasive exit strategies in DEXs. Attackers split sell volumes, spread actions across wallets, and bypass existing impact-/identity-based detectors via various combinations. Our model generalizes static behavioral thresholds into dynamic ones to capture such long-running exits.

We conducted a large-scale measurement of 303,614 on-chain liquidity pools and identified 105,434 scams. We also observed that multi-wallet exits have steadily replaced single-wallet drains, and direct owner participation has dropped from 65\% to 24\%. Modern scammers are becoming “smarter”. We issue a strong warning to DeFi communities.

%\newpage

\section*{Ethical Considerations}

Our study completely complies with ethical guidelines for computer security and information systems research, as outlined in the Menlo Report (listed by IEEE S\&P). All experiments and analysis were conducted on publicly available/accessible blockchain on-chain data from the public Ethereum blockchain, which is accessible for everyone on internet. Our measurement and experiment not contain interaction with human subjects, collection of personal identifiable information, or off-chain tracing of private individual/action by any form. All wallet addresses, transaction hashes, and smart contracts analyzed are public knowledge (if full-address and project name mentioned) or anonymized (by hiding the middle part of every involved wallet/token address involved in the study). 

Complying with responsible disclosure principles, we did not exploit or interact with any live DeFi protocol or smart contract beyond passive observation and public data collection to perform our experiment upon. We make no attempt to deanonymize, link, indicate or attribute them to real-world identities. No vulnerabilities or exploitable software flaws were disclosed during this research. 

We believe this work contributes positively to ecosystem security by empirically identifying evasive scam behaviors that undermine trust in decentralized finance. By making only aggregated and anonymized findings public, we mitigate any potential misuse of our detection logic while providing the research community with a transparent, reproducible framework for future scam analysis.

\section*{Open Science.} 

We release all the key materials/artifact that are used in this study to promote science transparency and support reproducibility to any future research within the same area. We public all the research-relevant dataset (anonymized), code and instruction, which can be accessible at~\cite{frpsource}. We encourage the community to reuse our artifacts, independently verify our measurements, and extend our research toward stronger defenses for DeFi ecosystems.

\section*{Ethical Usage of LLMs.} 

This work involved limited use of large language models (e.g., ChatGPT) solely for language refinement purposes. The models were used to improve grammar and typos. The content was written entirely by the authors. No parts of the manuscript, including conceptual content, technical descriptions, data analysis, or results, were generated by any GenAI or LLMs. All scientific ideas, analyses, and conclusions are the sole responsibility of the authors. 

\bibliographystyle{unsrt}
\bibliography{bib}

%\appendix
%\input{sec/11_appendix}

%the raw sharepoint folders:

%figure: https://rmiteduau-my.sharepoint.com/personal/s4037841_student_rmit_edu_au/_layouts/15/onedrive.aspx?id=%2Fpersonal%2Fs4037841%5Fstudent%5Frmit%5Fedu%5Fau%2FDocuments%2FRQ2%2DMaterial&ct=1762849659886&or=Teams%2DHL&ga=1&LOF=1

\end{document}